\documentstyle[aps,prl,epsfig]{revtex}

\def\cl{\centerline}
\def\bs{\bigskip}

\def\begt{\begin{tabular}}
\def\endt{\end{tabular}}
\def\bege{\begin{equation}}
\def\ende{\end{equation}}

\def\al{\alpha}
\def\be{\beta}
\def\ga{\gamma}
\def\de{\delta}
\def\eps{\epsilon}

\begin{document}

\title{Exactly solvable statistical model for two-way traffic}
\author{Vladislav Popkov 
\footnote{also at the Institute for Low Temperature Physics,
Kharkov, Ukraine.\\
E-mail: popkov@physik.fu-berlin.de} and Ingo Peschel\\
\small{Institut f\"ur Theoretische Physik, Freie Universit\"at Berlin,\\
Arnimallee 14, 14195 Berlin Germany}
}
\maketitle

\begin{abstract}
We generalize a recently introduced traffic model, where the statistical 
weights are associated with whole trajectories, to the case of two-way 
flow.
An interaction between the two lanes is included which describes a slowing down when two cars meet. This  leads to two coupled five-vertex models. 
It is shown that this problem can be solved by reducing it to two one-lane
 problems with modified parameters.
In contrast to stochastic models,
jamming appears only for very strong interaction between the lanes. 
\end{abstract}

\section{Introduction}

       The non-equilibrium properties of one-dimensional lattice gases
     have been studied intensively over the last years \cite{Zia}. With lattice
     gases, one can model not only physical situations such as transport
     in solid ionic conductors \cite{Peschel} or growth processes \cite{Krug},
     but also
     the traffic flow on roads \cite{Nagel}. Moreover, they can be used to study
     general features of phase transitions in non-equilibrium systems
     \cite{NATO}-\cite{Popkov_Schuetz}.
     For the traffic problem, the simplest model is the completely
     asymmetric exclusion process (ASEP), where classical hard-core
     particles hop stochastically, with unit rate, in one direction only 
     \cite{Gunter_habilitat}.
     On a ring, one then finds a steady state of product form where all
     configurations are equally likely. In terms of the density $\rho$ of 
     particles, the flux is then given by $j = \rho(1-\rho)$ and shows already the
     qualitative features found also in more sophisticated models, i.e.
     it vanishes for $\rho = 0,1$ and has a maximum in between.

       An essentially new description of traffic flow was proposed recently 
     by Brankov et al. \cite{Brankov}.
     In this work, non-intersecting domain-wall lines 
     on a square lattice were interpreted as space-time trajectories of cars.
     The weight of a trajectory is then obtained from the fugacities for
     horizontal and vertical moves. The single step, however, has no
     stochastic interpretation. The problem can be formulated in terms
     of a five-vertex model which generates these lines and which is
     exactly solvable since it satisfies the so-called free-fermion condition.
     The result for the flux $j$ is physically reasonable and very 
     similar to that for a variant of the (stochastic) Nagel-Schreckenberg
     model \cite{NS}.
         In this paper, we show that one can generalize this model to the
     case of two-way traffic  where cars on different lanes interact
     with each other.
The specific effect which we are treating is a tendency to slow down
when another car is approaching. In the two-dimensional formulation,
this is described by a modification of the fugacities whenever
trajectories of oppositely-moving cars cross.
 One then is led to consider two five-vertex models with
     a certain coupling between them. It turns out, however, that this
     coupling only renormalizes the parameters in each subsystem, so that
     the problem remains solvable as before. One finds that, in this model,
     the effect of an obstacle, i.e. of a car in the other lane,
 is relatively weak.
     While in stochastic models already a certain finite reduction of the 
     hopping rate at one position usually leads to a traffic-jam phenomenon
      with a region of high density appearing in front of the bottleneck
      \cite{Janowsky}-\cite{Hinrichsen}, this happens here only if the
     fugacity is reduced to
     zero for a large system.  As will be explained, this feature is related
     to the different weighting of the trajectories in both cases.
     
     In the following, we first describe the model in section 2 and then 
     explain its solution in section 3. Finally, in section 4, we discuss the
     results and add some further remarks.

\section{ Model} 

     We first recall the formulation of the original  one-way
     traffic model in \cite{Brankov}.  For a square lattice with periodic
     boundary conditions, the horizontal direction is interpreted
      as space, the vertical one as time (increasing
     downwards). Non-intersecting lines running towards 
     the lower right are then drawn on the lattice and viewed
     as trajectories of right-moving cars. They do not end,
     so that the number $N_1$ of cars is conserved. A horizontal
     step, representing a move, is given fugacity (weight)
     $x_1$, a vertical one fugacity $t_1$. Statistical averages are 
     then obtained from the partition function
     
\bege
Z( N_1,x_1,t_1) = \sum_C {x_1}^{N_{ x}(C)} {t_1}^{N_{ t}(C)} 
\label{Z_local}
\ende 
where $N_{ x}(C)$ and  $N_{ t}(C)$ are the  total numbers of 
     steps in the two directions for a certain configuration $C$ 
     of trajectories. 
        These trajectories are generated with their correct
      weights if at each lattice site the vertices shown in
      Figure~\ref{right_movers} are possible.
\begin{figure}

\setlength{\unitlength}{0.6truemm}
\begin{center}
\begin{picture}(160,50)(-35,-25)            
\thinlines

\linethickness{0.5mm}
\put(-20,0){\line(1,0){20}}
\put(40,0){\line(1,0){20}}
\thinlines
\put(10,0){\line(1,0){20}}
\put(70,0){\line(1,0){20}}
\put(100,0){\line(1,0){20}}
\put(130,0){\line(1,0){20}}
\linethickness{0.5mm}
\put(100,0){\line(1,0){10}}
\put(140,0){\line(1,0){10}}

\linethickness{0.5mm}
\put(-10,-10){\line(0,1){20}}
\put(-9,-15){1}
\thinlines
\put(20,-10){\line(0,1){20}}
\put(21,-15){2}
\linethickness{0.5mm}
\put(51,-15){3}
\put(80,-10){\line(0,1){20}}
\thinlines
\put(50,-10){\line(0,1){20}}
\put(81,-15){4}

\put(110,-10){\line(0,1){20}}
\put(111,-15){5}
\put(141,-15){6}
\put(140,-10){\line(0,1){20}}
\linethickness{0.5mm}
\put(110,-10){\line(0,1){10}}
\put(140,0){\line(0,1){10}}

\end{picture}
\end{center}
\caption{ Vertex configurations for  right-moving cars:
Boltzmann weights are $ 0,1,x_1,t_1,\sqrt{x_1t_1}$
respectively
}
\label{right_movers}
\end{figure}

        Since crossings (vertex 1) are forbidden, one is effectively
         dealing with a five-vertex model which can be
         solved exactly via the Bethe ansatz, even for more general
          weights $w_5$, $w_6$ \cite{Noh},\cite{Huang et al}. In the 
         present case, free-fermion techniques can be used to
         obtain the partition function \cite{Felderhof}.

           For the two-way traffic model, we introduce a second
         lattice where trajectories run towards the lower left,
         corresponding to the cars in the other lane. The fugacities
          are taken to be $x_2$, $t_2$ and the trajectories are now 
         generated by the vertices in Figure~\ref{left_movers}.

\begin{figure}

\setlength{\unitlength}{0.6truemm}
\begin{center}
\begin{picture}(160,50)(-35,-25)            
\thinlines

\linethickness{0.5mm}
\put(-20,0){\line(1,0){20}}
\put(40,0){\line(1,0){20}}
\thinlines
\put(10,0){\line(1,0){20}}
\put(70,0){\line(1,0){20}}
\put(100,0){\line(1,0){20}}
\put(130,0){\line(1,0){20}}
\linethickness{0.5mm}
\put(110,0){\line(1,0){10}}
\put(130,0){\line(1,0){10}}

\linethickness{0.5mm}
\put(-10,-10){\line(0,1){20}}
\put(-9,-15){1}
\thinlines
\put(20,-10){\line(0,1){20}}
\put(21,-15){2}
\linethickness{0.5mm}
\put(51,-15){3}
\put(80,-10){\line(0,1){20}}
\thinlines
\put(50,-10){\line(0,1){20}}
\put(81,-15){4}

\put(110,-10){\line(0,1){20}}
\put(140,-10){\line(0,1){20}}
\linethickness{0.5mm}
\put(110,-10){\line(0,1){10}}
\put(140,0){\line(0,1){10}}
\put(-9,-15){1}
\put(21,-15){2}
\put(51,-15){3}
\put(81,-15){4}
\put(111,-15){5}
\put(141,-15){6}
\end{picture}
\end{center}
\caption{ Vertex configurations for  left-moving cars:
Boltzmann weights are $ 0,1,x_2,t_2,\sqrt{x_2t_2}$
respectively
}
\label{left_movers}
\end{figure}

          To formulate the interaction between cars in the two
          lanes, the indices at the vertices are specified in the following
          way

\setlength{\unitlength}{0.6truemm}
\begin{center}
\begin{picture}(160,30)(-35,-15)            
\thinlines

\put(-20,0){\line(1,0){20}}
\put(70,0){\line(1,0){20}}
\put(-55,-3){lattice 1}
\put(100,-3){lattice 2}
\put(-20,2){$\al$}
\put(0,2){$\al'$}
\put(70,2){$\ga'$}
\put(90,2){$\ga$}

\put(-10,-10){\line(0,1){20}}
\put(80,-10){\line(0,1){20}}
\put(-9,10){$\be$}
\put(-9,-10){$\be'$}
\put(81,10){$\delta$}
\put(81,-10){$\delta'$}

\end{picture}
\end{center}
     The variables $\alpha, \beta,...$  take the value 1 if a car is present
             (thick line) and zero otherwise. For all vertices, the so-called 
            ice rule

 \bege
\al+\be = \al'+\be' \ \ \mbox{and}  \ \  \ga+\de = \ga'+\de'
\label{ice_rule}
\ende
holds, which ensures the conservation law for the number
            of cars,separately for both lanes.
 
              We now imagine that the two lattices are placed above each
            other and attribute an additional Boltzmann weight

\bege
v = exp(-\epsilon) = 
exp
\left(
{-h \over 2} 
\left(
\al \de + \al' \de' + \be \ga' + \be' \ga 
\right)
\right)
\label{eps_chosen}
\ende
to adjacent vertices in the two layers. Then each crossing of
             two trajectories will be weighed with the factor
\bege            
            0 < r = exp(-h) < 1                                       
\ende                  
  To see this, one first notes that $\epsilon = 0, v = 1$ if one of the 
vertices is of type 2, i.e. if there is no car present. The values of
  $\epsilon$ in the remaining cases are given in Table 1.
It then follows that simple crossings, which involve a pair of
      vertices of type 3 and 4, lead directly to a factor $r$,
see Table 1. 

\bs
\cl{Table 1. Interaction $\epsilon$ between two adjacent vertices in the
                              two layers.}

\begin{tabular}  {@{}lllll} 
$ vertex $ & $3$ & $4$ & $5$ & $6$ \\ \hline
$3$ & $0$ &$-h$ &$-h/2$  & $-h/2$ \\ 
$4$ & $-h$   & $0$ &$-h/2$  & $-h/2$ \\ 
$5$ & $-h/2$  & $-h/2$ &$-h/2$  & $-h/2$ \\    
$6$ & $-h/2$  & $-h/2$ &$-h/2$  & $-h/2$ \\      
\end{tabular}

\begin{figure}
\setlength{\unitlength}{0.47truemm}
\begin{center}
\begin{picture}(160,180)(-35,-15)
            
\thinlines


\put(-10,0){\line(1,0){160}}
\put(-10,20){\line(1,0){160}}
\put(-10,40){\line(1,0){160}}
\put(-10,60){\line(1,0){160}}
\put(-10,80){\line(1,0){160}}
\put(-10,100){\line(1,0){160}}
\put(-10,120){\line(1,0){160}}
\put(-10,140){\line(1,0){160}}


\put(0,-10){\line(0,1){160}}
\put(20,-10){\line(0,1){160}}
\put(40,-10){\line(0,1){160}}
\put(60,-10){\line(0,1){160}}
\put(80,-10){\line(0,1){160}}
\put(100,-10){\line(0,1){160}}
\put(120,-10){\line(0,1){160}}
\put(140,-10){\line(0,1){160}}


\put(30,170){\vector(1,0){60}}
\put(110,168){$x$}

\put(-30,110){\vector(0,-1){60}}
\put(-30,30){$t$}


\linethickness{0.5mm}



\put(100,-10){\line(0,1){10}}
\put(100,0){\line(0,1){20}}

\put(20,20){\line(0,1){20}}
\put(20,40){\line(0,1){20}}
\put(20,60){\line(0,1){20}}
\put(0,80){\line(0,1){20}}
\put(0,100){\line(0,1){20}}
\put(0,120){\line(0,1){20}}


\put(-10.4,140){\line(1,0){10.8}}
\put(-0.4,80){\line(1,0){20.8}}
\put(19.6,20){\line(1,0){20.8}}
\put(39.6,20){\line(1,0){20.8}}
\put(59.6,20){\line(1,0){20.8}}
\put(79.6,20){\line(1,0){20.8}}



\put(120,20){\line(0,1){20}}
\put(80,40){\line(0,1){20}}
\put(80,60){\line(0,1){20}}
\put(40,80){\line(0,1){20}}
\put(20,100){\line(0,1){20}}
\put(20,120){\line(0,1){20}}
\put(20,140){\line(0,1){10}}


\put(19.6,100){\line(1,0){20.8}}
\put(39.6,80){\line(1,0){20.8}}
\put(59.6,80){\line(1,0){20.8}}
\put(79.6,40){\line(1,0){20.8}}
\put(99.6,40){\line(1,0){20.8}}
\put(119.6,20){\line(1,0){20.8}}
\put(139.6,20){\line(1,0){10.8}}



\put(40,-10){\line(0,1){10}}
\put(40,0){\line(0,1){20}}
\put(40,20){\line(0,1){20}}
\put(80,40){\line(0,1){20}}
\put(100,80){\line(0,1){20}}
\put(140,100){\line(0,1){20}}
\put(140,120){\line(0,1){20}}
\put(140,140){\line(0,1){10}}


\put(39.6,40){\line(1,0){20.8}}
\put(59.6,40){\line(1,0){20.8}}
\put(79.6,80){\line(1,0){20.8}}
\put(99.6,100){\line(1,0){20.8}}
\put(119.6,100){\line(1,0){20.8}}



\put(60,140){\line(0,1){10}}
\put(60,120){\line(0,1){20}}
\put(80,100){\line(0,1){20}}
\put(120,80){\line(0,1){20}}
\put(140,60){\line(0,1){20}}

\put(59.6,120){\line(1,0){20.8}}
\put(79.6,100){\line(1,0){20.8}}
\put(119.6,80){\line(1,0){20.8}}
\put(139.6,60){\line(1,0){10.8}}


\put(-30,155){\small particle 1}
\put(10,155){\small particle 2}
\put(125,155){\small particle 4}
\put(50,155){\small particle 3}

\thinlines
\put(40,20){\circle{9.0}}
\put(80,60){\oval(8.0,48)}
\put(110,100){\oval(28,7.0)}

\end{picture}
\end{center}

\caption{ Trajectories of one 
left-moving and three right-moving particles with different types of crossings.}
\label{crossings}
\end{figure}
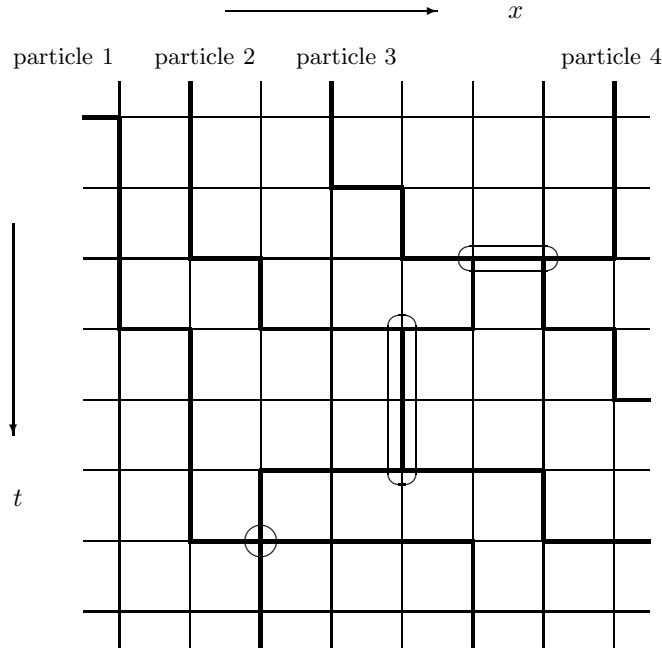

\bs     
 If trajectories meet
      and run (anti)parallel before they separate again, each of the
      two branch points contributes a factor $\sqrt{r}$. Some examples
      illustrating such crossings are shown in Figure~\ref{crossings}.

One should mention that the choice (\ref{eps_chosen}) 
       for the interaction
       is not unique. The more general form for $\epsilon$
                 
\bege
- \eps= A(\al \de'+\be\ga)+B(\al \de+\be'\ga) +C(\al' \de'+\be\ga')
+D(\al' \de+\be'\ga')
\label{eps_general}
\ende
      with $A+B+C+D = h$, still leads to the same factor $r=exp(-h)$.
      The individual terms listed in Table 1, however, become more
      complicated.

     In the model defined in this way, one still has the freedom to
     choose the particle numbers and the fugacities. Thus, by 
     setting $x_2 = 0$, one can immobilize the cars in the second lane and
     treat in particular the case of one fixed obstacle, which is of
     special interest.

\section{Solution} 

   We now show that the two-lane model can be solved by
  reducing it to the one-lane problem. The proof follows Ref.
  \cite{Huang} where a similar problem was treated. It is based
  on the ice rule (\ref{ice_rule}) which relates horizontal and vertical bond
  variables. Suppose that the lattices have $N$ columns and $M$ 
  rows, and let $\alpha_{n,m} (\gamma_{n,m})$ and $\beta_{n,m} (\delta
  _{n,m})$ be the variables to the right and below the vertex $(n,m)$,
  respectively, in the two layers. Then the total interaction is
$$
E=
- {h \over 2} \sum_{m=1}^M \sum_{n=1}^N 
(
\al_{ n-1,m}\de_{n,m-1} + \al_{ n,m}\de_{n,m} + \be_{n,m-1}\ga_{n-1,m} +
 \be_{n,m}\ga_{n,m} )
$$
  With the help of (\ref{ice_rule}), this can be rewritten as
  
\bege
 E = - {h \over 2} \left(
\sum_{m=1}^M (\al_{ 0,m} +  \al_{ N,m}) N_2 + 
\sum_{m=1}^M (\ga_{0,m} +  \ga_{N,m}) N_1 + U(0)-U(M)
\right)
\label{E}
\ende
where 
\bege
U(m)=\sum_{n=1}^{N-1} \sum_{k=1}^n 
(\be_{k,m} \de_{n+1,m} - \be_{n+1,m}\de_{k,m})
\ende
 contains only vertical bonds, while the other two terms in (\ref{E})
 contain only horizontal bonds. Due to the periodic boundary 
 conditions, the difference $U(0)-U(M)$ vanishes and one obtains
\bege
 E = -h N_2 \sum_{m=1}^M \al_{ N,m}  -h N_1 \sum_{m=1}^M \ga_{N,m} 
 \label{E_int}
\ende
 This can be compared with the effect of a rescaling  $x_1 \rightarrow x_1 e^{-\eta_1}$,
$x_2 \rightarrow x_2 e^{-\eta_2}$  which leads to an extra factor
 $exp(-\eta_1(\al + \al')/2 -\eta_2(\ga + \ga')/2 )$
for each pair of adjacent vertices.  Summed over all
  sites, this gives

$$
 E'= -\sum_{m=1}^M \sum_{n=1}^N \left( {\eta_1 \over 2}
 (\al_{ n-1,m} + \al_{ n,m}) + 
  {\eta_2 \over 2}
 (\ga_{n,m} + \ga_{n-1,m}) 
\right)
$$
  which can be expressed as
$$
 E'= -\sum_{m=1}^M  {\eta_1 \over 2}N (\al_{ 0,m} +  \al_{ N,m}) 
+V(0)-V(M) 
-\sum_{m=1}^M  {\eta_2 \over 2}N (\ga_{0,m} +  \ga_{N,m}) 
-\tilde V(0)+ \tilde V(M)
$$
      where 
\bege
V(m)=\sum_{n=1}^{N-1} \sum_{k=1}^n 
\be_{k,m} -
\sum_{n=2}^{N} \sum_{k=n}^N 
\be_{k,m} 
\ende
  and $\tilde V(m)$  is defined analogously with $\beta \rightarrow \delta$. Using 
  again the periodic boundary conditions, one finds
  
\bege
E' = -\eta_1 N \sum_{m=1}^M \al_{ N,m} 
-\eta_2 N \sum_{m=1}^M \ga_{N,m} 
\ende
 which has the same form as $E$ in (\ref{E_int}).
 Therefore the interaction 
 has the same effect as a change in the horizontal fugacities 
 if one chooses $\eta_1 = h N_2/N = h \rho_2$
 and $\eta_2 = h N_1/N = h \rho_1$
 where $\rho_1$ and $\rho_2$ are the densities of cars in the two lanes.
 The partition function is then
   
\bege
Z(N_1,N_2,x_1,x_2,t_1,t_2,r) = Z(N_1,x_1 r^{\rho_2},t_1)  
Z(N_2,x_2 r^{\rho_1},t_2) 
\label{Z_result}
\ende
    This exact formula looks like the result of a mean-field treatment
    since only the densities in the other layer enter the expressions.
    One should point out that it also holds for more general choices
    of the vertex weights in the layers. Then, also the weight $w_1$ of
    vertex 1 has to be renormalized with the same exponential factor.

\section{Results and discussion}

   One can now make use of the results for the 
  single-lane case \cite{Brankov}. For one lane, the 
  flux per site is equal to the average number of 
  horizontal steps and given by

\bege
j(\rho,x) = { \langle N_{ x}  \rangle \over NM}= { 1 \over 2}
\left[
 {1 \over \pi} 
 arccos
\left(
{ c - 2 x + c x^2
\over 
1 - 2 x c + x^2}
-1
\right) - \rho
\right]
\label{j}
\ende

  \noindent where $c = cos(\pi \rho)$ and $x < 1$ has been assumed.
  This is the physical region since the average speed
  of one car is $v = x/(1-x)$. As a function of $\rho$, 
  the flux has a maximum at $\rho = (1/{\pi})arccos(x)$, which
  shifts from $\rho = 1/2$ to $\rho = 0$ as $x$ increases.
  
    By inserting $x_1 r^{\rho_2}$ and $x_2 r^{\rho_1}$ into (14), one 
   then obtains the fluxes $j_1$ and $j_2$ in the two-lane case.
   These do not depend on the motion in the other lane,
   but only on the density there. Since $j$ increases with
   $x$, the interaction factor $r^{\rho_\alpha}$ always reduces the 
   flux, as expected. This reduction, however, becomes
   smaller as the density in the second lane decreases.
   For the case of only one car one has

\bege
j_1 = j(\rho_1, x_1 r^{1/N})
\ende
   and this approaches the value $j(\rho_1,x_1)$ without 
   interaction for large $N$. In order to slow down the 
   traffic appreciably, one would need $h \approx N$, i.e. an
   interaction increasing with the size, so that $r$
   vanishes exponentially. In other words, a transition
   occurs only at $r = 0$.
   
     As mentioned, the situation is different for stochastic
   models. There $j$ shows a sudden decrease as soon as the 
   corresponding quantity $r$ (describing the reduced crossing
   probability at a defect) falls below a certain finite value
   $r_c$. This is connected with the appearance of a jam at the
   defect. In terms of trajectories, the effect can be 
   described as follows. Consider a stochastic model as in
   \cite{Brankov} where a particle can move an arbitrary
   distance horizontally, at each step continuing with 
   probability $p$ and stopping with probability $q = 1-p$. At
   the defect, the quantities are $p'< p$ and $q'> q$. A particle
   some steps away from the defect will typically move to the
   bottleneck and then stay there for some time. Due to $q'> q$,
   such a trajectory has a higher weight than any other one
   where it makes stops before and then crosses the defect
   immediately. The same holds for another particle following
   it, since this has $q'= 1$ once it has reached the site next 
   to the first one. In this way, the jam builds as a region
   of vertical trajectories to the left of the defect.
   
     In the present model, the picture is different. There is
   no advantage in staying at the blockage, the crossing factor
   $r$ and the weights $x^k t^l$ are the same as for paths which
   approach the defect gradually. Nor is there an advantage for
   following particles to move next to the preceding one. 
   Therefore no jam builds up. One could say that the model 
   mimics the anticipation of disturbances by producing less
   densely packed trajectories. But, in shifting $r_c$ to zero, it
   overestimates the effect.
   
     It is also interesting to compare the two models at the
   operator level. According to \cite{Krinsky}, the transfer
   matrix $T$ of the (one-layer) five-vertex model commutes with 
   the operator
   
\bege
{\cal{H}}= - \sum_n \left( 
\sigma_n^x \sigma_{n+1}^x + \sigma_n^y \sigma_{n+1}^y + 2 H
\sigma_n^z\right)
\ende

   \noindent where $H = (1+x^2-t^2)/2x$, and it is easy to see that the
   ground state of $\cal{H}$ gives the maximal eigenvalue of $T$. This
operator 
   shows very clearly the free-fermion character of the model and
   also its non-stochastic nature, since the necessary $\sigma^z
   \sigma^z$-terms (which are related to the loss processes in the
   master equation) are missing.
   
   If one uses more general vertex weights
   $w_5$ and $w_6$, the operator

\bege
{\cal{H}}= - \sum_n \left( 
\sigma_n^- \sigma_{n+1}^+ + \Delta \sigma_n^z \sigma_{n+1}^z 
\right)
\ende
commutes with $T$, where $\Delta = (w_3w_4-w_5w_6)/(w_2 w_4)$ \cite{McCoy},
   \cite{Noh}. Although this contains such terms and has the form
   of the time-evolution operator for fully asymmetric hopping
   \cite{Gwa},\cite{Alcaraz}, the fact that $\Delta$ is not equal to
   one still makes it different. On the other hand, this model is
   interesting, because it contains, in the $x$--$t$ plane, a frozen phase
   with density $\rho = 1/2$  \cite{Noh},\cite{Huang et al}, where the 
   trajectories have the form of
   stairs with steps of unit lenght in both directions. This corresponds
   to synchronized traffic with always one empty site between the cars.
As this phase gives the highest possible throughput of vehicles 
and persists for a wide range of parameters $x,t$, it represents the
analogue of the maximal current phase in  stochastic hopping models
\cite{Krug91,Popkov_Schuetz}. 
In the $j$--$\rho$ relation, one then finds a cusp at $\rho = 1/2$.
   As mentioned above, also this model can be treated in the two-way
   case. However, apart from half-filling, the blocking properties
   will be similar to those described above.
   
\section*{Acknowledgement}
V.P. would like to thank the Alexander von Humboldt 
foundation for financial support.


\end{document}